\documentclass[aip,reprint]{revtex4-1}
\usepackage{lipsum}
\usepackage{graphicx}
\usepackage{dcolumn}
\usepackage{bm}
\usepackage[mathlines]{lineno}
\usepackage{hyperref}
\usepackage{xcolor}
\hypersetup{
    colorlinks=true,
    linkcolor=blue,
    citecolor=blue,
    urlcolor=blue,
}
\usepackage{color,soul}

\usepackage[T1]{fontenc}
\usepackage{siunitx}
\DeclareSIUnit\torr{Torr}
\DeclareSIUnit\oersted{Oe}
\draft 

\begin{document}



\title{Spin wave excitations in a nanowire spin-torque oscillator with perpendicular magnetic anisotropy}


\author{M. Succar}
\author{M. Haidar}

\email[Corresponding author: ]{mh280@aub.edu.lb}
\affiliation{Department of Physics, American University of Beirut, P.O. Box 11-0236, Riad El-Solh, Beirut 1107-2020, Lebanon}


\date{\today}

\begin{abstract}
Spin torque oscillators (STOs) are emerging microwave devices that can potentially be used in spin-logic devices and the next-generation high-speed computing architecture. Thanks to their non-linear nature, STOs are easily tunable by the magnetic field and the dc current.
Spin Hall nano-oscillators (SHNOs) are promising types of STOs and most of the current studies focus on localized modes that can be easily excited.  Here, we study using micromagnetic simulations, the nature of the spin-torque-induced excitations in nanowire devices made of perpendicular magnetic anisotropy (PMA) material. Our results showed that upon including PMA the excitation of localized and propagating spin wave modes is feasible. We study the nature of the mode excitations as a function of the PMA strength (\text{K}$_u$), and the current.
\hspace{0.1cm}Indeed, we estimate a critical value of \text{K}$_u$ to allow for the excitation of the propagating spin wave. We attribute this mode selectivity between localized and propagating modes to a switch in the sign of the nonlinearity of the system from negative to positive at a non-zero \text{K}$_u$ which is supported by analytical calculations. Our results provide deep insight into engineering reconfigurable microwave devices for future magnonic and computational applications.

\end{abstract}


\maketitle 

\section{Introduction}
Recently, the spin-orbit torque (SOT) generated by the spin Hall effect (SHE)\cite{D'yakonov1971,Hirsh1999} is a promising mechanism to excite, manipulate and detect magnetization dynamics in magnetic nanodevices\cite{Berger,Slonczewski}. 
Spin torque oscillators \cite{ DemidovSHNO,DemidovNCSHNO, Mohseni, Ranjbar2014,Awad2016,Collet2016,Haidar2019,Haowen2022} are an example of the emerging microwave devices which have a significant advantage in the field of magnonics, and can potentially be used in the next-generation high-speed computing architecture \cite{Locatelli2014,Zahedinejad2020}. Of particular interest for developing a high-speed signal processing paradigm utilizing spin waves are the excitation and the control of the propagating wave in STOs \cite{Houshang2018}. It is well demonstrated that SOT can drive self-localized excitations efficiently in different device layout  \cite{Liu2013,Duan2014,Giordano2014,Dvornik2018}. The nature of the excited mode is determined by the sign of the non-linearity coefficient, N, that depends on the orinetation of the magnetic field\cite{slavin2009}: (i) For in-plane fields \text{N} is negative (\text{N}<0), and a self-localized mode is excited where the frequency of the localized mode lies below the linear mode, i.e., the ferromagnetic resonance frequency (FMR). (ii) For out-of-plane fields \text{N} is positive (\text{N}>0) where a propagating spin wave can be driven at frequencies higher than FMR. A wide range of SHNOs of different geometry were studied where either the active magnetic layer is modified\cite{Liu2015,Divinskiy2017,Haidar2021} or heavy metals with very large spin Hall angle\cite{Liu2012,Kelller2019,Hamid2018} were used. In all these devices self-localized modes are reported.
Moreover, in PMA-based SHNOs devices, propagating spin waves are observed: In Bi-substituted YIG-based devices\cite{Evelt2018} with sufficiently large PMA, the nonlinear frequency shift vanishes leading to the current-independent frequency of auto-oscillations. On the contrary, a non-monotonic frequency shift is measured in metallic-based devices made of W/CoFeB/MgO nanoconstrictions \cite{Fulara2019}. While in both studies a propagating spin-wave is excited, the frequency shift is different which raises the question of how the characteristics of the propagating mode depend on PMA strength. A systematic study examining the role of PMA in devices is still lacking. 

In this paper, we perform micromagnetic simulations to study the SOT-driven excitation in a nanowire-based device made of CoFeB (3 nm)/Pt (5 nm) bilayers. We determine the nature of the excited modes whether localized or propagating as we vary the PMA strength (\text{K}$_u$), and the magnitude of dc current (\text{I}). Our results showed that in the nanowire device the excitation of a propagating spin wave is feasible upon including PMA in addition to the localized mode. Indeed, we estimate a critical value of \text{K}$_u$ to allow for the excitation of propagating spin waves. We attribute the mode transition between localized and propagating modes to a switch in the sign of the nonlinearity of the system from negative to positive at a non-zero \text{K}$_u$ which is supported by analytical calculations. This study demonstrates the role of PMA in determining the type of oscillations in nanodevices and helps in designing SHNOs devices with controlled characteristics.

\section{Micromagnetic simulations}

We simulate a stack of 3 nm CoFeB and 5 nm Pt layers containing a rectangular-shaped nanowire of 50 nm width and 200 nm length Fig. 1(a). The electrical current density and the corresponding Oersted field in the devices were simulated using COMSOL software under a reference electrical current of \text{I}$_{ref}$ = \SI{2}{mA} 
 as shown in Figs. 1(b, c). Since both metallic layers conduct electrical currents, the calculations show that the electrical current flows mostly in the Pt layer due to the difference in the resistance between the CoFeB and Pt layers.  
We perform micromagnetic simulations using mumax$^{3}$ solver with the input from the COMSOL simulation. In the simulations we assume a rectangular mesh that matches the COMSOL mesh. The mesh has dimension of \SI{2000}{nm}$\times$\SI{2000}{nm}$\times$\SI{3}{nm} with a cell size of \SI{3.9}{nm}$\times$\SI{3.9}{nm}$\times$\SI{3}{nm}. The spin current density (\text{J}$_{s}$), is calculated from the electrical current density (\text{J}$_{e}$) using the relation $J_\text{s} = \theta_\text{SH} J_\text{e}$  where $\theta_\text{SH}$ is the spin Hall angle of Pt and is equal to \SI{0.1}{} . For micromagnetic simulations, we assume the CoFeB/Pt bilayers have a saturation magnetization of $\mu_{0} M_\text{s}$ = \SI{0.9}{T}, a Gilbert damping $\alpha$ of 0.02, a gyromagnetic ratio $\gamma/2\pi$ of \SI{30}{GHz/T} and an exchange stiffness of \SI{10}{pJ/m}, consistent with experimental studies. 
The magnetic field, $\mu_{0}$H, is applied at a fixed in-plane angle of \ang{20} and oriented at \ang{75} out-of-plane with a constant value of \SI{0.7}{T}. The perpendicular magnetic anisotropy is varied between \SI{0}{MJ/m^3} to \SI{0.2}{MJ/m^3}. 
The magnetization dynamics is simulated by integrating the Landau-Lifshits-Gilbert-Slonczewski equation over \SI{250}{ns}. The frequencies and spatial profiles of excited modes of the system are extracted by performing the Fast Fourier Transform of the time domain data that represents the evolution of the magnetization.
\begin{figure}[!h]
\begin{center}
\includegraphics[width=0.48\textwidth]{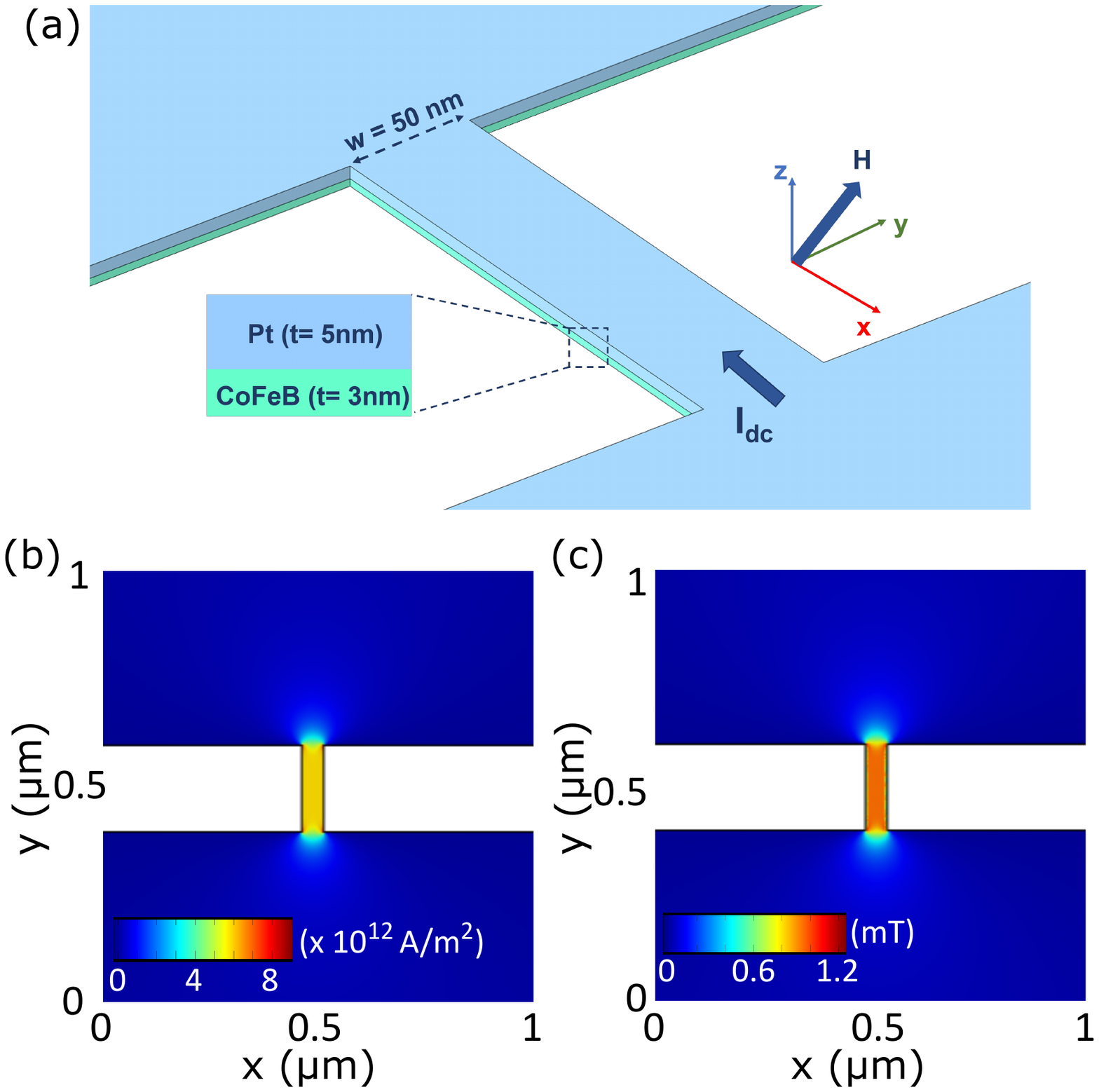}
\caption{\label{fig1} (Color online) (a) Schematic layout of the nanowire-based spin-torque oscillator device made of  CoFeB (3 nm)/Pt (5 nm) bilayer with a nanowire width of \SI{50}{nm}. This shows the direction of the applied magnetic field and the dc current. (b, c) Current density and Oersted field distribution along x- and y-direction respectively of the nanowire device calculated at I = 2 mA using COMSOL software.}.
\end{center}
\end{figure}

\section{Results and Discussion}

\begin{figure}[!h]
\begin{center}
\includegraphics[width=0.34\textwidth]{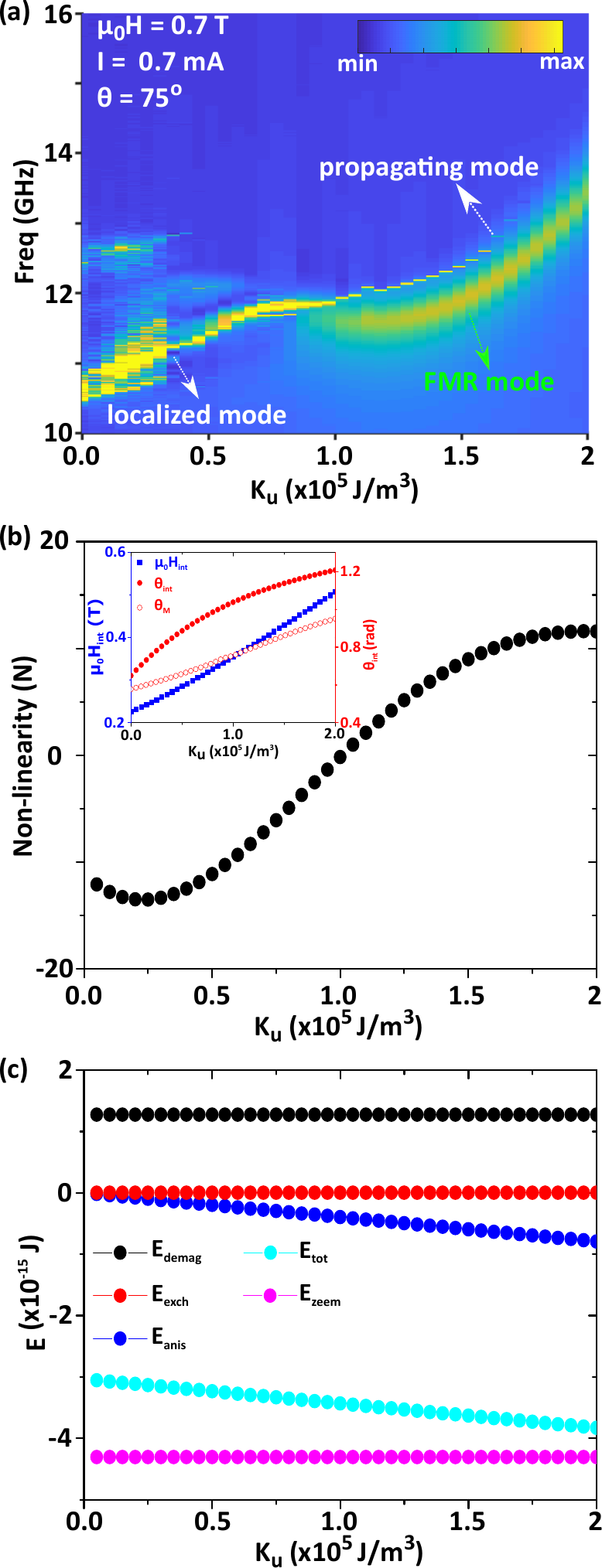}
\caption{\label{fig2} (Color online) (a) Power spectral density of spin wave excitations in CoFeB/Pt nanowire as a function of \text{K}$_{u}$. The nanowire is magnetized out-of-plane at \ang{75} under a magnetic field of magnitude $\mu_0 H = 0.7$ T, an in-plane dc current of magnitude \text{I} = \SI{0.7}{mA} is applied. (b) The variation of the non-linearity coefficient is calculated numerically as a function of \text{K}$_{u}$. (inset) The variation of the effective internal magnetic field (blue color), angle (red closed dots), and the magnetization out-of-plane angle (red open dots) extracted from the micromagnetic simulation as a function of \text{K}$_{u}$. (c) Simulated energy of the ground magnetization state as a function of \text{K}$_{u}$.}.
\end{center}
\end{figure}
The nanowire is magnetized out-of-plane by applying a magnetic field of \SI{0.7}{T} at \ang{75} and a current of \SI{0.7}{mA} is injected in-plane to look into the excited spin wave modes by the spin-torque. Figure 2(a) shows the power spectral density calculated while changing the magnitude of \text{K}$_u$. One can observe two clear modes with different signatures. A low frequency mode which is characterized by a wide linewidth and a low amplitude and is excited even at zero PMA value. The dispersion of this mode matches the FMR frequency, hence it is associated to the FMR mode of the devices. The increase in the frequency of the FMR mode at higher \text{K}$_u$ is due to a larger effective magnetic field. In addition, another mode is excited where it is characterized by a sharp peak with a high amplitude which is the signature of current-induced auto-oscillations. We attribute this mode to SOT driven spin-wave auto-oscillations. Note that for low- \text{K}$_u$ the frequency of the auto-oscillations lies below the FMR mode which is the signature of a localized mode excitation. As the magnitude of \text{K}$_u$ reaches above a critical value of \SI{0.1}{MJ/m^3}, the frequency of the oscillations are detected above the FMR frequency which hints that a propagating spin wave is excited in the nanowire device. To confirm this, we calculate analytically the non-linearity coefficient (\text{N}) of the nanowire using Eq. (1) in the Supplementary material. The inset of Fig. 2(b) shows the variation of the internal field ($\mu_{0}$\text{H}$_{int}$) and the internal angle ($\theta_{int}$) extracted from the micromagnetic simulations to accurately calculate \text{N} for a nanowire and the magnetization out-of-plane angle ($\theta_{M}$) as a function of \text{K}$_u$.
The main set of Fig. 2(b) shows the variation of \text{N} versus \text{K}$_u$. We calculate a negative \text{N} at \text{K}$_u$ = 0, it further decreases forming a minimum for low-\text{K}$_u$, then it increases as \text{K}$_u$ increases conserving a negative sign for \text{K}$_u$ below \SI{0.1}{MJ/m^3}. Since \text{N} is negative in this regime, only localized mode can be excited \cite{Bonetti2010}. In our simulations, this mode can be observed at lower current values as we will discuss below. As we increase \text{K}$_u$ above \SI{0.1}{MJ/m^3}, \text{N} switches its sign from negative to positive and hence it promotes the excitation of propagating spin-wave excitations. This result showed that one can manipulate the non-linearity by modifying PMA strength and hence control the characteristics of the excited mode between a localized or a propagating mode. Experimentally, the strength of PMA can be controlled in several ways such as engineering an oxide interface\cite{Demasius2016,Mondal2017,Zahedinejad2018apl}, tuning the PMA using ion irradiation\cite{Chappert1998,Herrera2015,Jiang2020}, and reducing the thickness of the active magnetic layer to the sub-nanometer range.     
To understand how PMA affects the nonlinearity, we look into the energy terms extracted from the micromagnetic sumulations. Figure 2(c) shows the variation of the  magnetic energies: the Zeeman energy (\text{E}$_{zeem}$), demagnetic energy (\text{E}$_{demag}$), anisotropy energy (\text{E}$_{K}$), exchange energy (\text{E}$_{ex}$), and total energy (\text{E}$_{tot}$) as a function of the PMA strength. Among these energies, \text{E}$_{demag}$ and \text{E}$_{anis}$ compete with each other as they have opposite signs. Note that while \text{E}$_{anis}$ becomes larger in negative as the PMA strength increases the \text{E}$_{demag}$ is not affected. The competition between those two factors defines the equilibrium direction of the magnetization such that the demagnetization energy prefers the magnetization to be in the film plane i.e. in this region where the non-linearity is negative; while the PMA pulls the magnetization out-of-plane i.e. to the region where the non-linearity is positive. In the present situation, the magnetization is oriented out-of-plane at \ang{35} for \text{K}$_{u} = $ \SI{0}{MJ/m^3} and it increases up to \ang{60} for \text{K}$_{u} = $ \SI{0.2}{MJ/m^3}. It is worth-noting that as \text{N} approaches zero at a critical PMA strength, or in other words a critical direction of the magnetization, a major improvement is expected in the linewidth of the auto-oscillation \cite{Rippard2006,Kim2008}.
\begin{figure*}
    \centering
\includegraphics[width=0.8\textwidth]{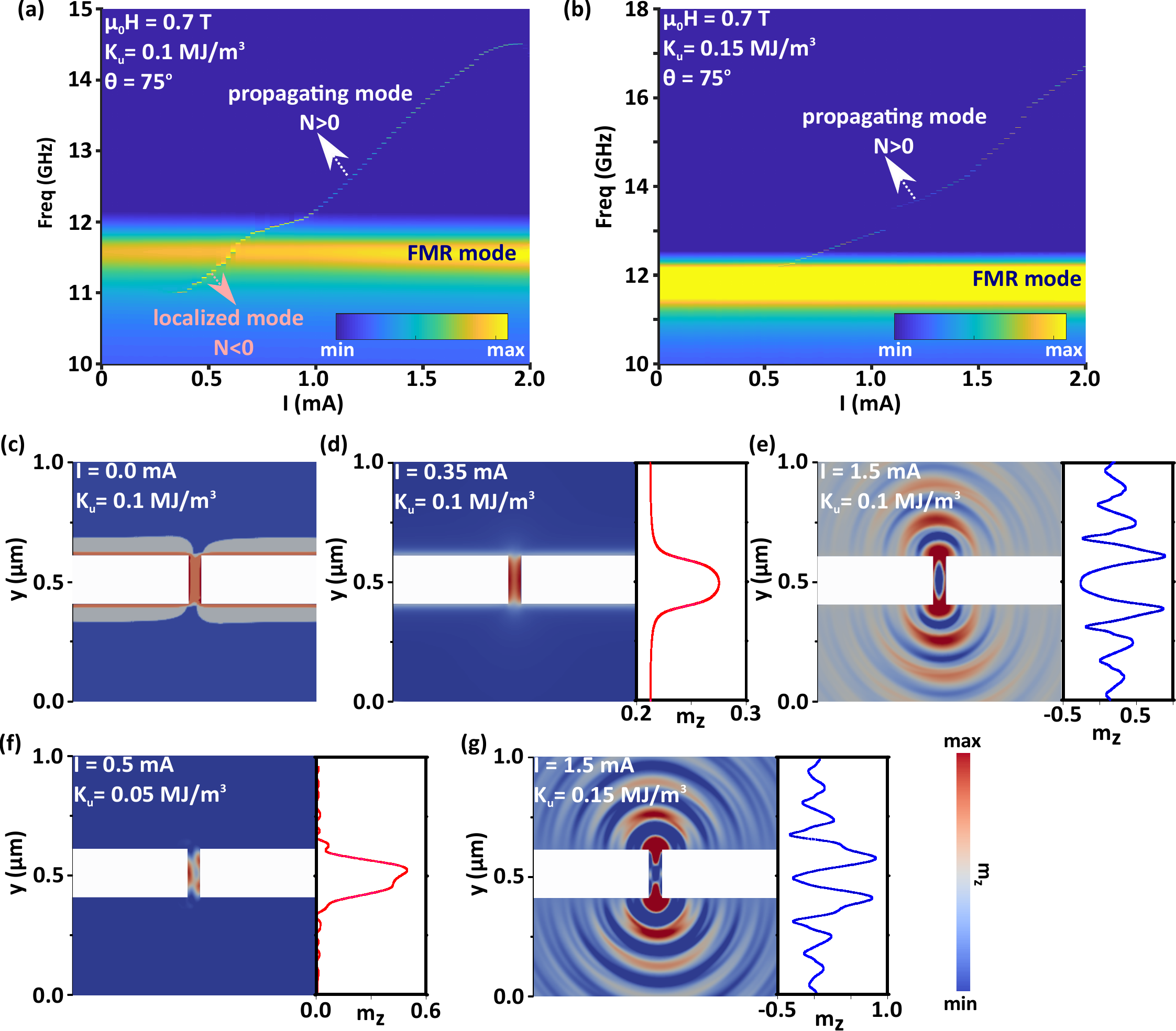}
\caption{\label{fig3} (Color online) Color plot of the power spectra of spin wave modes in CoFeB/Pt nanowire as a function of \text{I}$_{dc}$ at two values of \text{K}$_{u}$: (a) \SI{0.1}{MJ/m^3}, and (b) \SI{0.15}{MJ/m^3}. The calculated spatial mode profile of the z-component of the magnetization (\text{m}$_{z}$) at \textit{x} = \SI{0.5}{\mu m} of (c) the linear mode, and the auto-oscillations at (d) I = \SI{0.35}{mA}, and (e) I = \SI{1.5}{mA} with \text{K}$_{u} = $ \SI{0.1}{MJ/m^3}. The mode profiles are calculated at (f) low-\text{K}$_{u}$ = \SI{0.05}{MJ/m^3} and low-\text{I} = \SI{0.5}{mA} and (g) high-\text{K}$_{u}$ = \SI{0.15}{MJ/m^3} and high-\text{I} = \SI{1.5}{mA}. The micromagnetic simulations were carried under a magnetic field of magnitude $\mu_0 H$ = \SI{0.7}{T} and applied at out-of-plane \ang{75}.}
\end{figure*}

Next, we discuss how the dc current affects the characteristics of the excited modes. We consider now, the nanowire is magnetized out-of-plane by applying a magnetic field of \SI{0.7}{T} at \ang{75} and assuming two values of \text{K}$_{u}$: \SI{0.1}, and \SI{0.15}{MJ/m^3}.  We perform the simulations while scanning the current from \SI{0} to \SI{2}{mA}, and we calculate the power spectral density as shown in Fig 3(a--b). For \text{K}$_{u}<$ \SI{0.1}{MJ/m^3}, the non-linearity coefficient is negative and hence only localized mode can be excited as discussed above\cite{Bonetti2010}. Localized mode excitation is experimentally observed for in-plane magnetized nanowire\cite{Duan2014}. With the increase of \text{K}$_{u}$ to \SI{0.1}{MJ/m^3} three auto-oscillatory modes are observed. A current independent mode representing the FMR mode with a constant frequency \textit{f} = \SI{11.5}{GHz} is detected. The spatial mode profile of the linear FMR mode is shown in Fig. 3(c). At low currents, we observe oscillations starting from \SI{0.3}{mA} with a monotonic increase in the frequency of the auto-oscillations as we increase the current. The auto-oscillation in this regime appears below the FMR frequency. We calculate the spatial profile of the auto-oscillation at I = 0.35 mA as shown in Fig. 3(d). We note that the amplitude of the z-component of the magnetization (\text{m}$_{z}$) at \textit{x} = \SI{0.5}{\mu m} is localized within the vicinity of the nanowire while it is constant outside. For completeness, we calculate the mode profile at another combination for low-\text{K}$_{u}$ = \SI{0.05}{MJ/m^3} and low-\text{I} = \SI{0.5}{mA} regime as shown in Fig. 3(f). Indeed, we observe that the spatial profile of the auto-oscillation is confined within the region of the nanowire. Hence, We attribute the auto-oscillation in the low-\text{K}$_{u}$ and low-\text{I} regime to the excitation of a self-localized mode. 

At high currents (\text{I >} \SI{0.7}{mA}) a high-frequency mode is observed that oscillates at frequencies higher than the FMR frequencies, and it increases as the dc current increases and oscillates up to \textit{f =} \SI{14.8}{GHz}. The spatial profile of the auto-oscillation extracted at \text{I =} \SI{1.5}{mA} is shown in fig. 3(e). The amplitude of \text{m}$_{z}$ at \textit{x} = \SI{0.5}{\mu m} changes gradually between positive, zero and negative values where the oscillations extend over the device area. Note that the amplitude of the oscillation decays as the wave propagate far from the nanowire region. Therefore, the high current mode is attributed to a propagating spin-wave excitation. We estimate the wavelength of the propagating mode is \SI{140}{nm} in the nanowire. While considering a moderate \text{K}$_{u}$ = \SI{0.1}{MJ/m^3}, one can tune the sign of the non-linearity and hence select the excited mode between either a localized or a propagating spin-wave excitations by using a dc current. Interestingly, this mode transition is only possible for \text{N} near zero where the current contributes to determine the sign of \text{N} in a non-trivial way. This can be explained as following: At a critical \text{K}$_{u}$, in the present case \SI{0.1}{MJ/m^3}, the competition between the demagnetizing field and PMA results in a small negative \text{N}, while increasing the current the precessional angle of the magnetization increases which in its turn lowers the contribution of the demagnetizing field. As a result a new equilibrium is established, and hence the sign of \text{N} can switch from negative to positive value taking into account the current induced change to the demagnetizing field, a more quantitatively discussion can be found in \cite{Dvornik2018_2}. 
With a further increase of \text{K}$_{u}$ to \SI{0.15}{MJ/m^3}, as shown in Fig. 3(b) the FMR mode and a propagating spin wave mode are observed owing to \text{N}>0. Indeed, we can conclude that at high-\text{K}$_{u}$, the excitation of propagating spins is possible, as in our simulation this is supported by two factors: First, a blue shift in the frequency of the propagating mode is observed. Second, the amplitude of \text{m}$_{z}$ calculated at \textit{x} = \SI{0.5}{\mu m} oscillates over the nanowire with a reduced amplitude as shown in Fig. 3(g). Note that by increasing the \text{K}$_{u}$ spin waves with high frequencies can be achieved, up to \SI{20}{GHz} in the present study but could be pushed to higher values at different field orientation which could potentially be used in high-speed logic.

In conclusion, we study the impact of PMA strength on SOT-driven oscillations in CoFeB/Pt nanowire device. Our results showed that the sign of the non-linearity coefficient can be tuned from negative to positive by modifying the PMA strength and magnitude of the dc current. We showed both localized and propagating spin wave modes can be excited efficiently in the nanowire device. This study provides a deep insight for engineering reconfigurable microwave devices for various future applications in the field of magnonics and for new paradigms in neuromorphic computing.

See the supplementary material for the calculation of the non-linearity coefficient of the nanowire device.

This work was supported by the  American  University of Beirut Research Board (URB), the Mamdouha El-Sayed Bobst Deanship Fund, and the Faculty of Arts and Sciences Undergraduate Research Experience program (FAS-URE).

\textbf{DATA AVAILABILITY}

The data that supports the findings of this study are available from the corresponding author upon reasonable request.



%
\end{document}